\documentclass[sigconf, nonacm, pdfa]{acmart}
\usepackage[a-2b]{pdfx}

\usepackage{xspace}
\usepackage{xcolor}
\usepackage{amsmath}
\usepackage{amsfonts}
\usepackage{graphicx}
\usepackage{tabularx}
\usepackage{enumitem}
\usepackage{subcaption}

\newlist{compactitem}{itemize}{3} % 3 is max-depth
\setlist[compactitem]{label=\textbullet, nosep, leftmargin=0cm,itemindent=.5cm,listparindent=\parindent}

\newcommand\vldbdoi{10.14778/3611540.3611624}
\newcommand\vldbpages{4074 - 4077}
\newcommand\vldbvolume{16}
\newcommand\vldbissue{12}
\newcommand\vldbyear{2023}
\newcommand\vldbauthors{\authors}
\newcommand\vldbtitle{\shorttitle} 
\newcommand\vldbavailabilityurl{}
\newcommand\vldbpagestyle{empty}
\newcommand{\sys}{\textsc{Lingua Manga~}}

\begin{document}
\title{\sys: A Generic Large Language Model Centric System for Data Curation}

%%
%% The "author" command and its associated commands are used to define the authors and their affiliations.
\author{Zui Chen}
\affiliation{%
  \institution{IIIS, Tsinghua University}
  \city{Beijing}
  \country{China}
}
\email{chenzui19@mails.tsinghua.edu.cn}

\author{Lei Cao}
\affiliation{%
  \institution{U of Arizona/MIT}
  \city{Tucson}
  \state{Arizona}
}
\email{lcao@csail.mit.edu}

\author{Sam Madden}
\affiliation{%
  \institution{MIT}
  \city{Cambridge}
  \state{Massachusetts}
}
\email{madden@csail.mit.edu}

\begin{abstract}
Data curation is a wide-ranging area which contains many critical but time-consuming data processing tasks. However, the diversity of such tasks makes it challenging to develop a general-purpose data curation system. To address this issue, we present \sys, a user-friendly and versatile system that utilizes pre-trained large language models. \sys offers automatic optimization for achieving high performance and label efficiency while facilitating flexible and rapid development. Through three example applications with distinct objectives and users of varying levels of technical proficiency, we demonstrate that \sys can effectively assist both skilled programmers and low-code or even no-code users in addressing data curation challenges.
\end{abstract}

\maketitle

\pagestyle{\vldbpagestyle}
\begingroup\small\noindent\raggedright\textbf{PVLDB Reference Format:}\\
\vldbauthors. \vldbtitle. PVLDB, \vldbvolume(\vldbissue): \vldbpages, \vldbyear.\\
\href{https://doi.org/\vldbdoi}{doi:\vldbdoi}
\endgroup
\begingroup
\renewcommand\thefootnote{}\footnote{\noindent
This work is licensed under the Creative Commons BY-NC-ND 4.0 International License. Visit \url{https://creativecommons.org/licenses/by-nc-nd/4.0/} to view a copy of this license. For any use beyond those covered by this license, obtain permission by emailing \href{mailto:info@vldb.org}{info@vldb.org}. Copyright is held by the owner/author(s). Publication rights licensed to the VLDB Endowment. \\
\raggedright Proceedings of the VLDB Endowment, Vol. \vldbvolume, No. \vldbissue\ %
ISSN 2150-8097. \\
\href{https://doi.org/\vldbdoi}{doi:\vldbdoi} \\
}\addtocounter{footnote}{-1}\endgroup

\ifdefempty{\vldbavailabilityurl}{}{
\vspace{.3cm}
\begingroup\small\noindent\raggedright\textbf{PVLDB Artifact Availability:}\\
The source code, data, and/or other artifacts have been made available at \url{\vldbavailabilityurl}.
\endgroup
}

\section{Introduction}
\begin{sloppypar}
In the era of big data, organizations are collecting vast amounts of data from various sources. However, the data is often messy, incomplete, or contains errors. To effectively perform data analytics or any other applications, it is often necessary to engage in a data curation process~\cite{data-tamer,bigdc}, which includes tasks like data discovery, integration, and cleaning. Nonetheless, devising a data curation solution for a particular scenario can be a cumbersome and time-consuming process. It entails extensive communication of requirements, a collaboration between domain experts and programmers, multiple rounds of debugging and testing, and active maintenance to accommodate new use cases.
This underscores the need for a {\it generic} system that enables users to efficiently and effectively address diverse data curation challenges and apply solutions to various applications.

However, building a system that addresses diverse data curation problems and applications is challenging.
Data curation involves a range of distinct tasks, such as data discovery through table search, schema matching, entity resolution during data integration, and data imputation in data cleaning. Moreover, in real applications, various extra tasks are often involved in the data curation processes, such as event extraction, name extraction, anomaly detection, data summarization, visualization, table semantic parsing, etc. Worst yet, applications in different domains have diverse requirements. Effectively resolving data curation problems often relies heavily on users' domain knowledge, making developing a {\it general}-{\it purpose} solution improbable. For example, entity resolution over {\it books} focuses on text understanding, while entity resolution over {\it images} should rather pay special attention to image processing. As a result, existing data curation systems typically focus on only a few tasks on specific data formats. For example, Data Tamer~\cite{data-tamer} solves schema matching and entity resolution problems; ZenCrowd~\cite{zencrowd} targets entity linking; while CrowdDB~\cite{crowddb} addresses fuzzy SQL semantics.

Recent advances in Large Language Models (LLMs) have shed light on this highly challenging yet critical data curation area. Pre-trained LLMs such as GPT-3~\cite{gpt3}, Codex~\cite{codex}, ChatGPT~\cite{chatgpt}, and LLaMA~\cite{llama} have not only shown the ability to understand human needs and instructions but also to generate code like a programmer. Such ability could be leveraged to solve specific data curation problems like data imputation or entity resolution~\cite{fm}.

However, the current use of LLMs and LLM plugins is still far from the ultimate solution. Repeatedly calling LLM services can be costly and may lead to data privacy issues. For instance, consider a large e-commerce organization with thousands of documents and tables containing billions of rows -- it would be neither affordable nor secure to let an LLM access the entire data lake to complete the tasks. LLMs tend to be less effective without access to enterprise data due to the gap between the enterprise data and the public data LLMs trained on. Thus, there is an urgent need to harness the power of LLMs systematically, efficiently, and securely. Properly integrating LLMs and traditional data curation methods could be a promising path toward a general-purpose data curation system.

\begin{figure}[t]
  \centering
  \includegraphics[width=\linewidth]{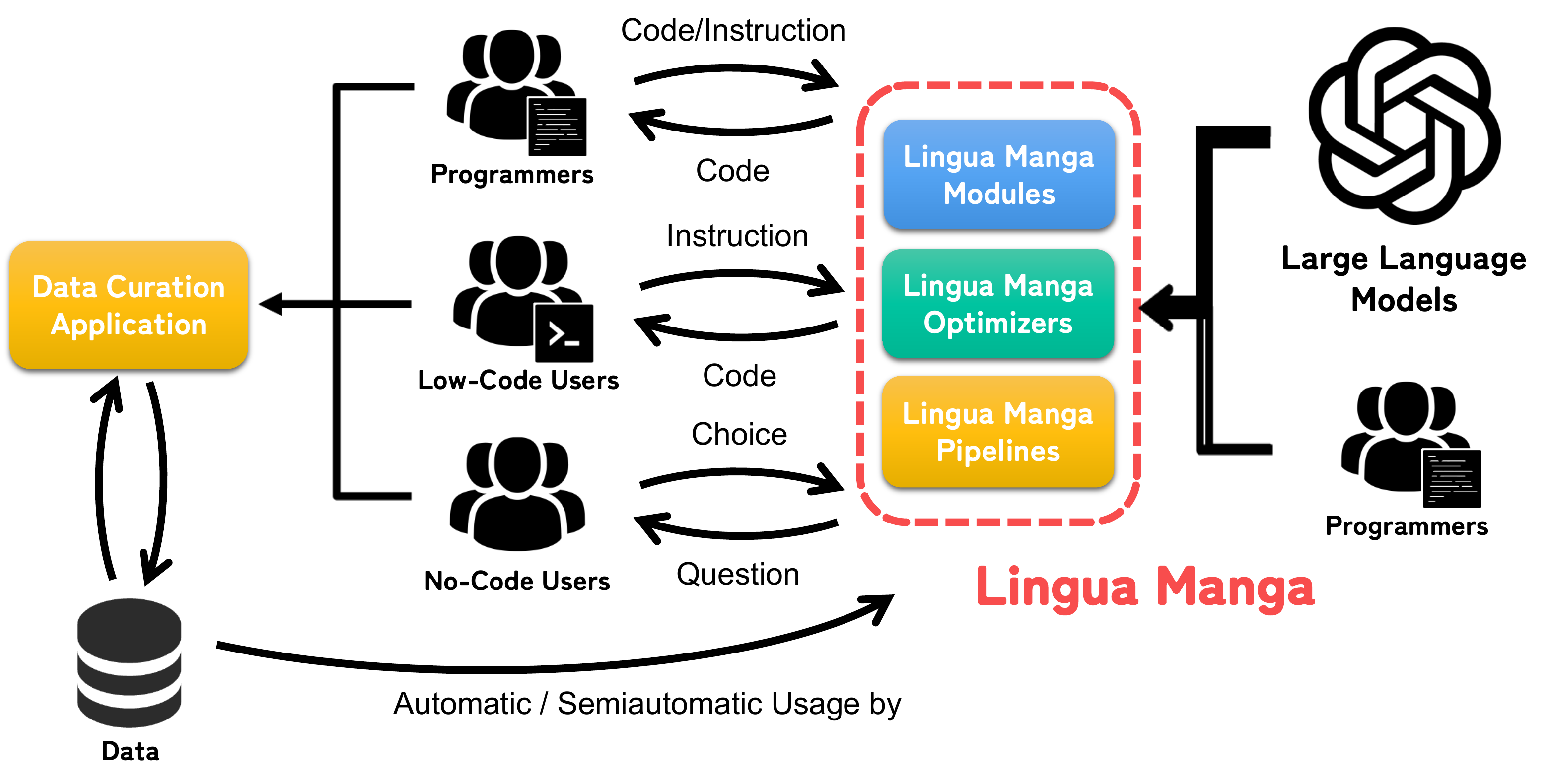}
  \vspace{-8mm}
  \caption{An illustration of the \sys System.}
  \label{fig:lingua-manga}
  \vspace{-8mm}
\end{figure}

In this demonstration, we present \sys, an LLM-centric system that enables the swift development of effective yet efficient data curation solutions. It benefits programmers as well as low-code or even no-code users. Specifically, \sys shows the following key properties:

\begin{compactitem}
    \item \textbf{User-friendly}: Users can quickly develop a data curation solution by utilizing the templates, built-in modules, and, most importantly, by effectively interacting with LLMs through \sys as a middleware.
    
	\item \textbf{Flexible}: Users can effortlessly communicate with \sys through natural language (NL) to inject domain knowledge or instructions into the data curation solution such that it can meet the needs of the specific applications or domains.
 
	\item \textbf{Intelligent}: \sys automatically optimizes the solution to fix errors, improves performance, continuously learns from the data, etc.
 
	\item \textbf{Highly Performant}: \sys minimizes the frequency of calling the LLM service, which incurs heavy computation and monetary costs, thus resulting in an economical, efficient, and privacy-preserving solution.

    \item \textbf{Label Efficient}: with our system, users can develop a data curation solution with no or only a few labeled examples from the specific application while still achieving accuracy comparable to the SOTA ML-based methods trained with thousands of labels.
\end{compactitem}

\end{sloppypar}
\section{Preliminaries: Prompt-tuning and Code Generation in LLMs}
\begin{sloppypar}

Starting from GPT-3~\cite{gpt3}, LLMs' size and computational cost have grown so much that users can no longer afford to fine-tune them for every downstream task. In the meantime, LLMs have also gained significant knowledge through training on large-scale corpora, leading to the emergence of a new paradigm known as prompt-tuning~\cite{prompt}. Using prompt-tuning, users can generalize LLMs to new tasks and data domains without modifying the LLM itself. This is achieved by incorporating prompts, such as descriptions, tags, and learnable components, into the input data.

Moreover, cutting-edge LLMs like ChatGPT~\cite{chatgpt} have demonstrated their effectiveness in interacting with humans using natural-language-based prompts \cite{awesome-chatgpt-prompts}. These prompts are interpretable by humans. For instance, a straightforward prompt like "Please determine if the following entities are equivalent" can enable LLMs to carry out entity resolution. This capability is especially useful in \sys since it facilitates LLMs to interactively execute functions vaguely specified by users in natural language.

Furthermore, prompts like ``Please write a code for entity resolution'' trigger LLMs to generate custom software modules via code generation. GPT-3~\cite{gpt3}, ChatGPT~\cite{chatgpt}, Codex, and Copilot~\cite{codex} are the mainstream LLMs that support code generation.
However, existing methods that use these models to generate code have limitations~\cite{nci}. They are typically inadequate for producing code with complex logic and building large-scale software. This calls for a new methodology to effectively utilize LLM Generated Code (LLMGC).

\end{sloppypar}
\section{\sys}
\begin{sloppypar}

\sys comprises a Domain-Specific Language (DSL), a Compiler, an Optimizer, and a set of templates.

\sys is a \textbf{user-friendly} workflow system that enables users to quickly build data curation solutions by composing pipelines of logical operators. The system features a DSL to simplify the workflow-building process.

\sys is \textbf{automatic}. Like a relational database, it auto-compiles each logical operator into a physical, executable {\it module}.

\sys is \textbf{extensible}. \sys allows the programmers to implement their own physical modules, which can be seamlessly plugged into the system as long as they follow the standard interfaces of \sys.

\sys is \textbf{flexible}. To meet the specific need of different applications, the \sys optimizer interacts with users via LLMs to generate customized physical modules. 

\sys is \textbf{intelligent}. The \sys optimizer uses LLMs to enhance the modules' capability.

In addition, to further facilitate users in building pipelines, \sys offers templates, essentially groups of pre-built pipelines. Rather than creating a pipeline from scratch, \sys allows users to start with a pre-defined, well-optimized pipeline that the target application can directly use.

The fundamental observation underlying this research is that while current LLMs may not yet be able to produce exceedingly intricate and large software systems, they are proficient at generating modular code on a smaller scale. Moreover, LLMs possess learning and logical reasoning ability that enable them to complete data curation tasks directly, indicating that LLMs could accelerate the process of developing a data curation solution by automating parts of it and, thus, reduce the workload of software developers.

Next, we discuss the modules and the \sys optimizer, which are the critical components of \sys.

\subsection{Modules in \sys} 

A module is a function $f:X \to Y$ that takes $X$ as input and returns $Y$. Modules are usually viewed as black boxes, and their behavior is determined by their expected output. In \sys, modules can be classified into four types:

\begin{compactitem}
    \item \textbf{Custom Module}: As a basic module, a custom module is implemented with manually written code. It can be created by users with programming skills or provided by \sys as a default built-in module.
    
    \item \textbf{LLM Module}: An LLM itself can be a module. An LLM module tends to be more powerful than a traditional module because of the common sense knowledge and logical inference abilities possessed by an LLM. However, an LLM module requires a good task description as input; and LLM outputs typically need proper validation, as textual responses generated by the LLMs could be diverse and unstable. For instance, numerical validation that ensures the LLM's output is within a reasonable value range can identify and correct potential errors in LLM output.
    
    \item \textbf{LLMGC Module}: An LLM can dynamically generate code to implement an LLMGC module, replacing the role of programmers. \sys allows LLMGC to call other modules in the system or use external tools such as a calculator or another pre-trained LLM. Although an LLMGC module might be less expressive than custom modules written by skilled programmers, it enables no-code or low-code users to implement or customize a module effortlessly.
    
    \item \textbf{Decorated Module}: As the most advanced module in \sys, a decorated module can comprise multiple basic modules and be enhanced by the optimizer that leverages the common sense knowledge encoded in the LLMs and their learning ability.
\end{compactitem}

\subsection{\sys Optimizer}

The optimizer in \sys can improve the performance of physical modules within a pipeline by enhancing their efficiency and effectiveness, e.g., avoiding repeatedly calling the LLM service to minimize the execution costs and the potential data privacy leak. Unlike conventional database optimizers, the optimizer in \sys is modularized and can be selectively composed by users to serve specific applications best. The key modules in \sys include:

%\subsubsection{Adaptor}

\begin{compactitem}
    \item \textbf{Validator}: It checks whether the target module behaves correctly on a few example test cases. It then uses the failed test cases to trigger the LLM to improve the target module and fix the errors. Specifically, the validator first calls an LLM to generate the suggestion by reading the code and the failure cases. Then, the code, failure cases, and the generated suggestion are sent to another LLM to generate a new version of the code. This validation cycle repeats until either all test cases are executed successfully, or a timeout ensues, leading to a re-generation of the LLMGC module until an additional timeout.

    \item \textbf{Simulator}: {A simulator automatically generates a more efficient and equally effective alternative to a given module that already functions well. For example, a module that frequently calls LLM service can be expensive. Potentially, a supervised learning counterpart could replace it. This is because LLMs are versatile, while a data curation task typically only needs part of its functionality. It is thus possible to produce a module proficient in simulating the specific LLM functionality required by a task.
    
    Because each module is treated as a black-box function, an ML-based simulator can replicate the target module through supervised learning. The target module will function as intended during initialization, and a control logic will decide when the simulated version should take over, such as after achieving the desired accuracy or reaching a certain level of confidence.
    
    A simulator could improve accuracy as well. At first sight, this is counter-intuitive. Like the teacher-student model in machine learning, traditionally, the student model, which learns from a teacher model, is expected to be upper bounded by the teacher model on performance. However, studies have shown that self-training with filters~\cite{bootstrapping,bootstrapping2,distillation,toolformer} can produce a superior student model because of better generalization. Moreover, the simulator will continuously monitor the real data flow in the original pipeline. It can thus constantly learn to adapt to the data distribution, potentially outperforming the original static module.}
    
    \item \textbf{Connector}: Concerning efficiency and data privacy, it is crucial for applications to reduce the amount of data exposed to LLMs, while still ensuring the effectiveness. 
    For instance, in an NL2SQL task, granting LLMs access to the entire table contents is neither secure nor cost-effective. However, relying solely on the table schema is not adequate to generate accurate predicates. To address this issue, a locally-running connector can be employed to manage the selective data upload to LLMs. A pre-defined connector for tabular data enables LLMs to execute SQL commands in local databases and obtain the resulting data while ensuring that the execution is limited to the queries specified by the user. In multi-modal data scenarios, appropriate pre-defined connectors are provided, such as connectors designed for handling extensive textual data.
\end{compactitem}

\end{sloppypar}
\section{Demonstration}
\begin{sloppypar}

\begin{figure}[t]
  \centering
  \includegraphics[width=0.95\linewidth]{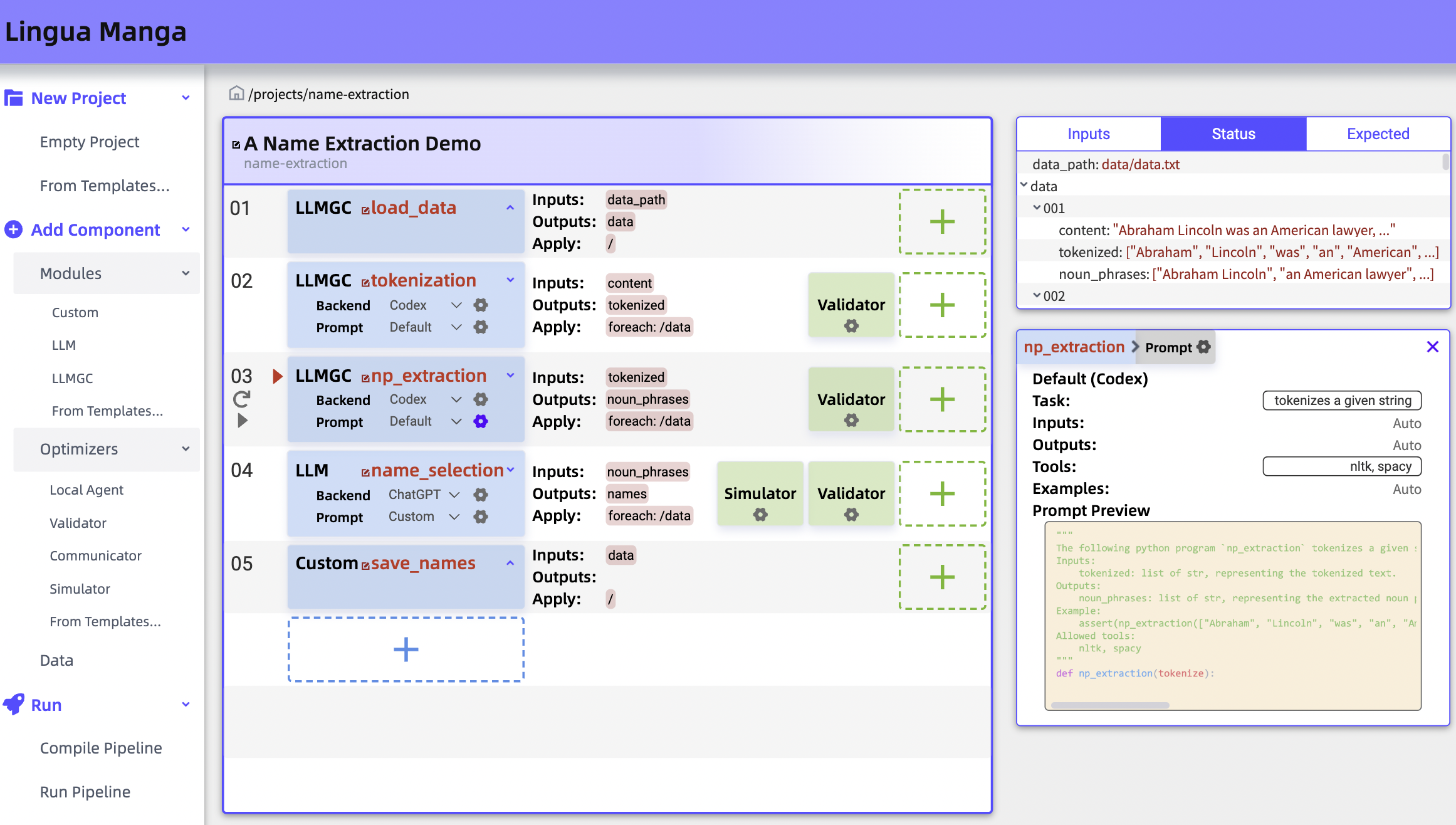}
  \vspace{-3mm}
  \caption{UI of the \sys System.}
  \label{fig:UI}
  \vspace{-6mm}
\end{figure}

We use three applications to demonstrate \sys. Fig.~\ref{fig:UI} shows its user interfaces with name extraction as an example task.

\vspace{-2mm}
  
\subsection{Entity Resolution: Effortless to the Novices}

Consider the scenario where a technical novice wants to perform entity resolution on a dataset but lacks programming experience.

\begin{figure}[t]
  \vspace{-4mm}
  \centering
  \begin{subfigure}{0.236\textwidth}
    \includegraphics[width=\linewidth]{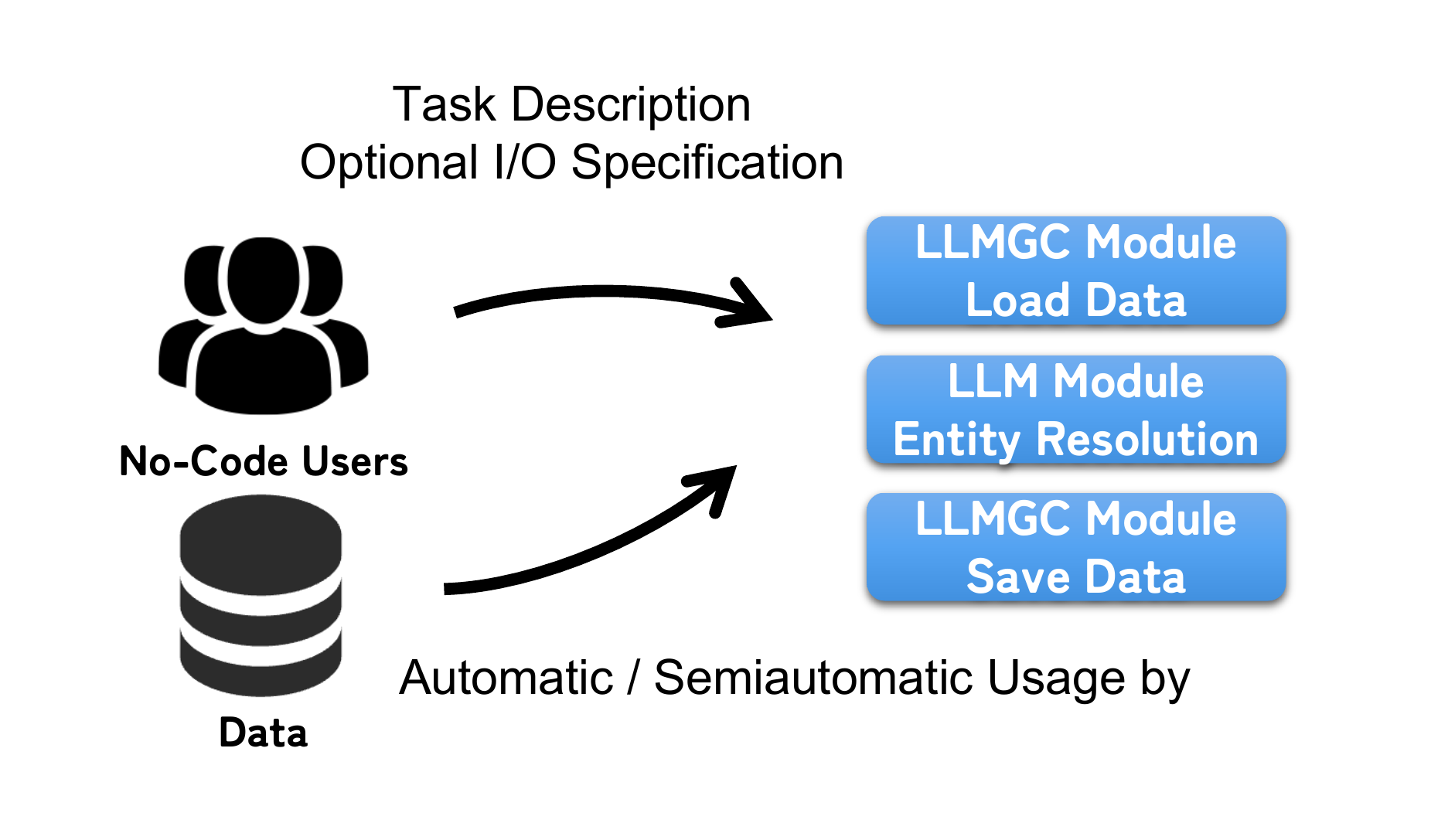}
    \caption{Custom pipeline.} \label{fig:E1a}
  \end{subfigure}%
  \hspace*{\fill}
  \begin{subfigure}{0.236\textwidth}
    \includegraphics[width=\linewidth]{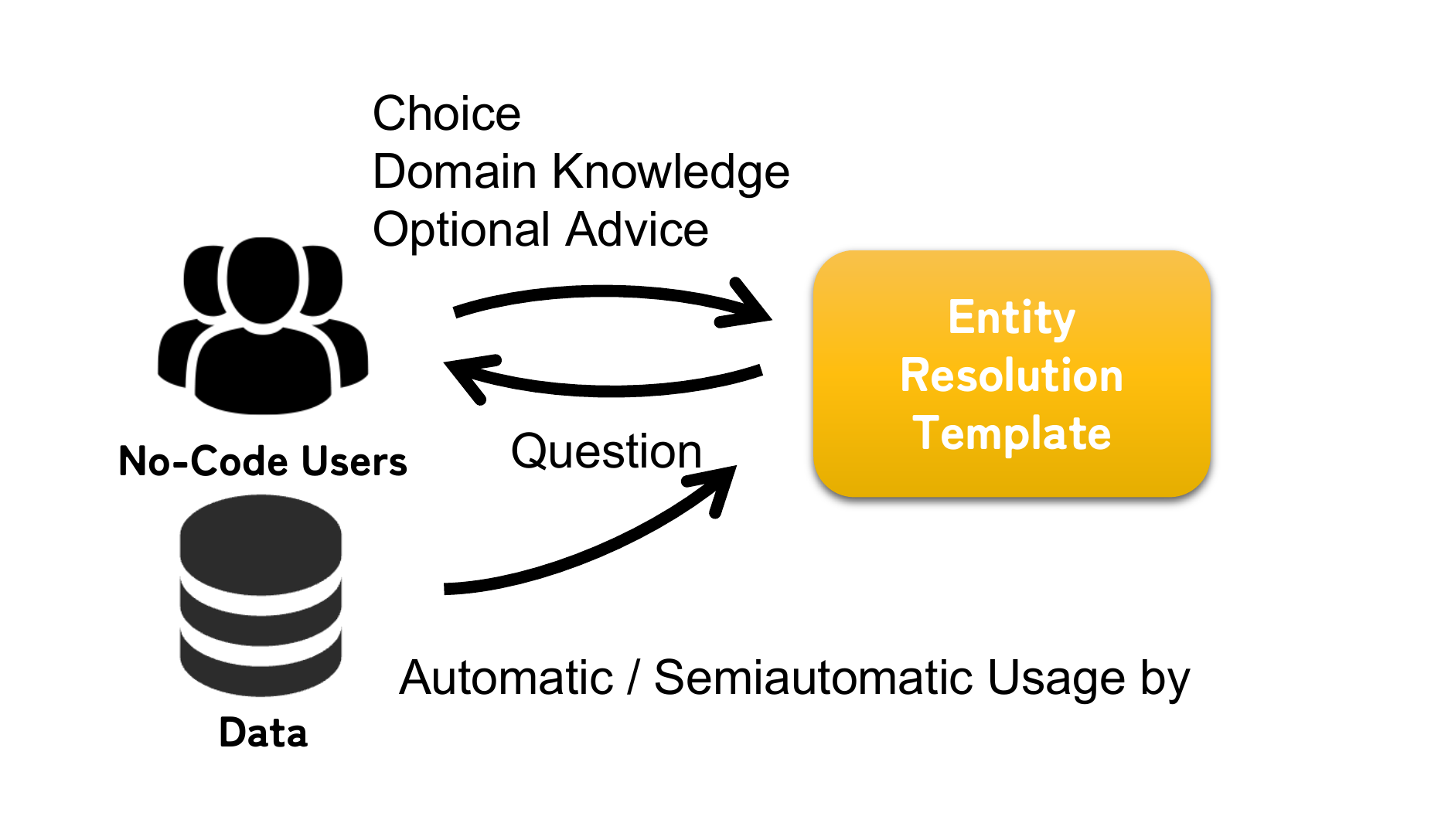}
    \caption{Builit-in template.} \label{fig:E1b}
  \end{subfigure}%
  \vspace{-4mm}
  \caption{An illustration of two possible Entity Resolution workflows using \sys.}
  \label{fig:E1}
  \vspace{-4mm}
\end{figure}

\sys effectively fulfills these requirements by providing a user-friendly solution. To begin, users can easily search for existing templates within the system. If none are available, they can create a simple pipeline that includes data loading, entity resolution, and data-saving operators, as depicted in Figure~\ref{fig:E1a}. It is worth noting that the user does not have to write any code. Instead, they can simply describe the task to the LLMs using the suggested prompt templates and provide optional input and output specifications through examples. The \sys optimizer will then improve the performance automatically.

We evaluated the F1 scores of the solution produced by \sys on three entity resolution datasets: BeerAdvo-RateBeer, Fodors-Zagats, and iTunes-Amazon~\cite{magellan}. As shown in Table~\ref{tab:E1}\footnote{The numbers of the baselines are obtained from~\cite{fm}.}, it significantly outperforms the previous LLM-based method~\cite{fm}, while achieving performance comparable to supervised learning algorithms trained with hundreds or even thousands of labeled examples~\cite{ditto}.

\begin{table}[htbp]
  \vspace{-2mm}
  \caption{Quantitative Experiment on Entity Resolution.}
  \vspace{-4mm}
  \label{tab:E1}
  \footnotesize
  \begin{tabularx}{0.96\linewidth}{lcccc}
    \toprule
    Dataset & Magellan~\cite{magellan} & Ditto~\cite{ditto} & FMs~\cite{fm} & \sys \\
    \midrule
    BeerAdvo-RateBeer & 78.8 & 94.37 & 78.6 & 89.66 \\
    Fodors-Zagats & 100.0 & 100.00 & 87.2 & 95.65 \\
    iTunes-Amazon & 91.2 & 97.06 & 65.9 & 92.00 \\
    \bottomrule
  \end{tabularx}
  \vspace{-4mm}
\end{table}

\subsection{Name Extraction: Flexible for the Adepts}

\begin{figure}[t]
  \centering
  \includegraphics[width=0.8\linewidth]{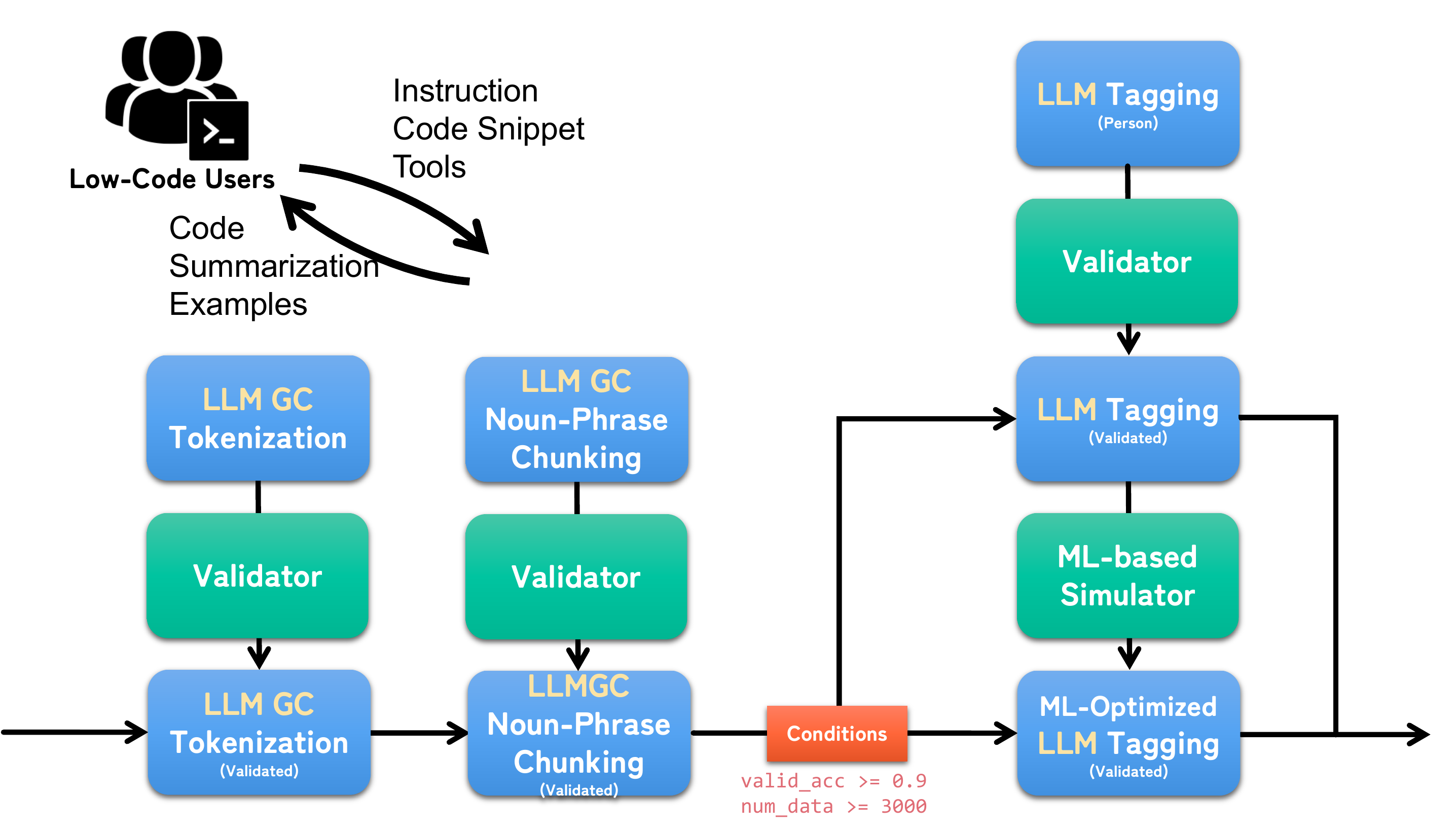}
  \vspace{-4mm}
  \caption{A possible Name Extraction pipeline.}
  \label{fig:E2}
  \vspace{-2mm}
\end{figure}

Consider a scenario where a domain expert, as a low-code user with basic programming skills, aims to get a high-accuracy and efficient name extraction solution, i.e., find all person names in a text passage. 
In this case, the domain expert understands that name extraction typically involves three steps: tokenization, noun-phrase extraction, and tagging. As a result, the domain expert can create a pipeline of three individual operators (ignoring data loading and saving for ease of presentation), as shown in Fig.~\ref{fig:E2}.

The domain expert recognizes that tagging is a complex process and is often the performance bottleneck. Therefore, he or she chooses to utilize the LLM module within \sys along with an example-based validator.
The other two operators are relatively simple and thus are realized as LLMGC modules -- using LLM to generate the code. Suppose the resulting LLMGC modules, such as the noun-phrase extraction module, are not precise enough. The domain expert can further enhance them by providing external tool APIs, domain knowledge instructions, or code snippets to optimize the code generation process.
Additionally, since repeatedly using LLM to tag each data record can be costly, the domain expert may use the simulator to create an ML-based alternative solution that simulates LLM tagging with significantly lower expenses.

To showcase the versatility of \sys, we experimented on a real-world name extraction dataset obtained from a startup company. This task is unique in that it has to handle multi-lingual data, which significantly degrades the accuracy of name extraction. \sys quickly resolves this issue by incorporating an LLM language detection module and providing multi-lingual tools to the LLMGC module. This shows that \sys can swiftly adapt to specific needs, being flexible and efficient.

\vspace{-1mm}
\subsection{Data Imputation: Excellency with the Experts}

Consider a scenario where an expert programmer needs to optimize a specific data imputation solution at all costs. This demo assumes the expert is dealing with the Buy dataset~\cite{imp,fm}, which comprises three attributes: products with names, descriptions, and manufacturers. The manufacturers' names are missing.

Because \sys effectively integrates LLMs, it is advantageous over traditional rule-based or learning-based approaches. For example, without having the relevant knowledge, it is almost impossible for the traditional methods to deduce that the product ``PlayStation 2 Memory Card 8MB'' is produced by ``Sony''. This is where the LLM module could help. 

\begin{figure}[t]
  \centering
  \vspace{-3mm}
  \includegraphics[width=0.40\linewidth]{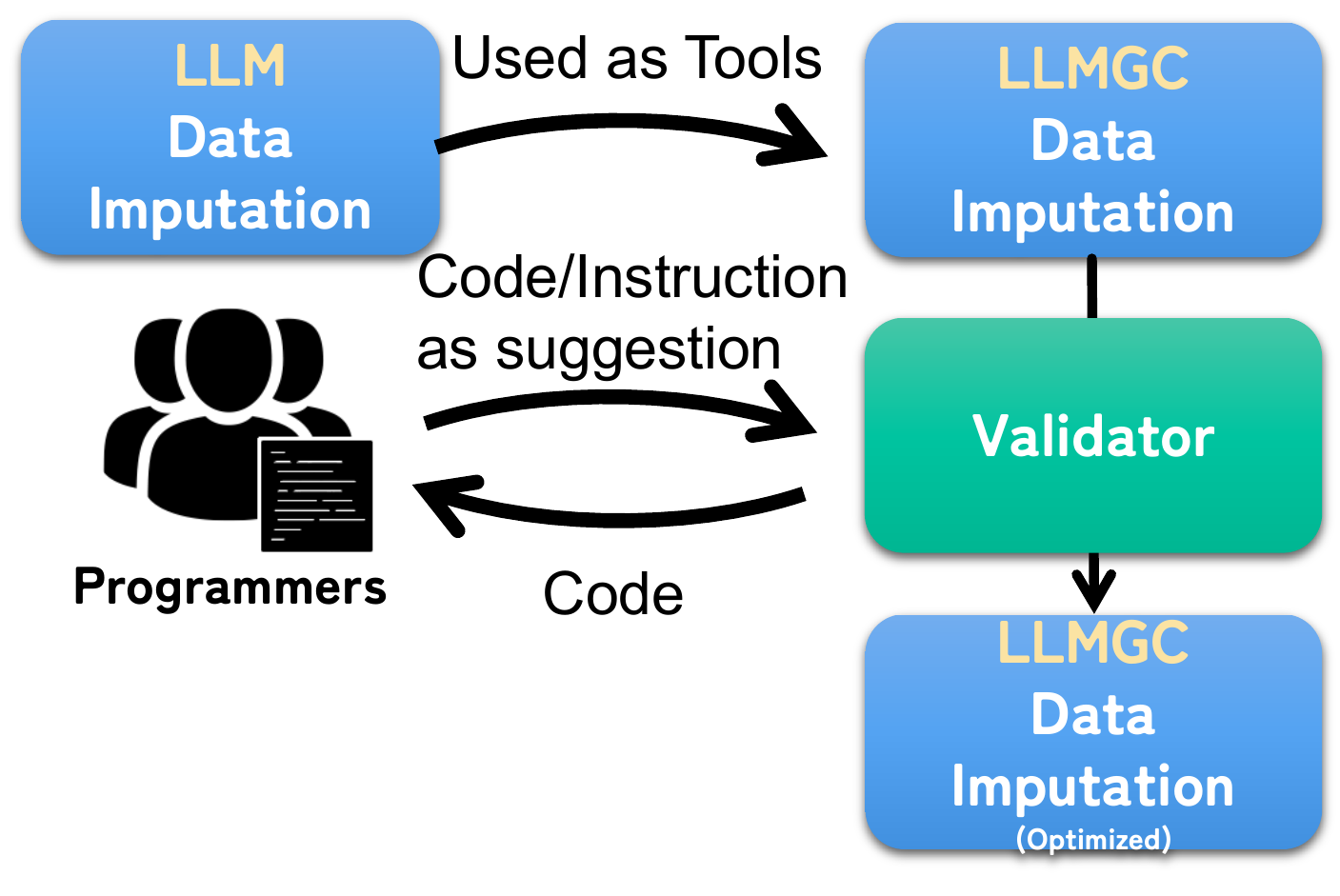}
  \vspace{-4mm}
  \caption{A possible Data Imputation pipeline.}
  \label{fig:E3}
  \vspace{-7mm}
\end{figure}

Experts can offer comprehensive guidelines or codes to create a better LLMGC module, as illustrated in Figure \ref{fig:E3}. These guidelines facilitate the validation procedure, enabling the LLMGC module to undergo iterative refinement. As a result, it can effectively use the LLM as an external tool to resolve complex cases while performing more efficiently than a pure LLM module on easy cases.

The \sys produced solution results in an accuracy of $94.48\%$, which is a remarkable improvement compared to the $16.2\%$ accuracy by HoloClean~\cite{holoclean}. It is comparable to IMP~\cite{imp}, which trains a Transformer model on thousands of training examples and achieves an accuracy of $96.5\%$.
It is noteworthy that our optimized version, which combines the LLM module with the LLMGC module, is cost-effective, using only $1/6$ LLM calls to achieve higher accuracy, compared to the version that only uses the LLM module to call LLM service repeatedly (accuracy $93.92\%$). Previous LLM-based research achieved an accuracy of $84.6\%$~\cite{fm}.

\end{sloppypar}
\vspace{-1mm}
\section{Conclusion}
\begin{sloppypar}

We introduced \sys, a user-friendly and versatile system designed to facilitate the development of data curation applications, leveraging the capabilities of LLMs. We demonstrated the efficiency, effectiveness, and user-friendliness of \sys through three data curation tasks. Nonetheless, it should be noted that \sys is currently in the prototype stage, which presents numerous opportunities for enhancements. These improvements encompass pipeline optimizations, supporting multi-modal applications, and implementing robust mitigation strategies to tackle LLM-induced hallucinations, etc.

\end{sloppypar}

%\clearpage

\vspace{-2mm}
\bibliographystyle{ACM-Reference-Format}
\bibliography{ref/ref-llm,ref/ref-prompt,ref/ref-curation,ref/ref-tool,ref/ref-exp}


\begin{thebibliography}{20}

%%% the Web address.

\ifx \showCODEN    \undefined \def \showCODEN     #1{\unskip}     \fi
\ifx \showDOI      \undefined \def \showDOI       #1{#1}\fi
\ifx \showISBNx    \undefined \def \showISBNx     #1{\unskip}     \fi
\ifx \showISBNxiii \undefined \def \showISBNxiii  #1{\unskip}     \fi
\ifx \showISSN     \undefined \def \showISSN      #1{\unskip}     \fi
\ifx \showLCCN     \undefined \def \showLCCN      #1{\unskip}     \fi
\ifx \shownote     \undefined \def \shownote      #1{#1}          \fi
\ifx \showarticletitle \undefined \def \showarticletitle #1{#1}   \fi
\ifx \showURL      \undefined \def \showURL       {\relax}        \fi
% invisible to TeX
\providecommand\bibfield[2]{#2}
\providecommand\bibinfo[2]{#2}
\providecommand\natexlab[1]{#1}
\providecommand\showeprint[2][]{arXiv:#2}

\bibitem[\protect\citeauthoryear{Das}{Das}{[n.d.]}]%
        {magellan}
\bibfield{author}{\bibinfo{person}{Sanjib et~al. Das}.}
  \bibinfo{year}{[n.d.]}\natexlab{}.
\newblock \bibinfo{title}{The Magellan Data Repository}.
\newblock
\newblock


\bibitem[\protect\citeauthoryear{et~al.}{et~al.}{2022}]%
        {fm}
\bibfield{author}{\bibinfo{person}{Avanika~Narayan et al.}}
  \bibinfo{year}{2022}\natexlab{}.
\newblock \showarticletitle{Can Foundation Models Wrangle Your Data?}
\newblock \bibinfo{journal}{\emph{{VLDB}}} (\bibinfo{year}{2022}).
\newblock


\bibitem[\protect\citeauthoryear{et~al.}{et~al.}{2023a}]%
        {awesome-chatgpt-prompts}
\bibfield{author}{\bibinfo{person}{Fatih Kadir~Akın et al.}}
  \bibinfo{year}{2023}\natexlab{a}.
\newblock \bibinfo{title}{Awesome ChatGPT Prompts}.
\newblock
\newblock


\bibitem[\protect\citeauthoryear{et~al.}{et~al.}{2012}]%
        {zencrowd}
\bibfield{author}{\bibinfo{person}{Gianluca~Demartini et al.}}
  \bibinfo{year}{2012}\natexlab{}.
\newblock \showarticletitle{ZenCrowd: leveraging probabilistic reasoning and
  crowdsourcing techniques for large-scale entity linking}. In
  \bibinfo{booktitle}{\emph{{WWW}}}.
\newblock


\bibitem[\protect\citeauthoryear{et~al.}{et~al.}{2023b}]%
        {llama}
\bibfield{author}{\bibinfo{person}{Hugo~Touvron et al.}}
  \bibinfo{year}{2023}\natexlab{b}.
\newblock \showarticletitle{LLaMA: Open and Efficient Foundation Language
  Models}.
\newblock  (\bibinfo{year}{2023}).
\newblock


\bibitem[\protect\citeauthoryear{et~al.}{et~al.}{2021a}]%
        {codex}
\bibfield{author}{\bibinfo{person}{Mark~Chen et al.}}
  \bibinfo{year}{2021}\natexlab{a}.
\newblock \showarticletitle{Evaluating Large Language Models Trained on Code}.
\newblock  (\bibinfo{year}{2021}).
\newblock


\bibitem[\protect\citeauthoryear{et~al.}{et~al.}{2011}]%
        {crowddb}
\bibfield{author}{\bibinfo{person}{Michael J.~Franklin et al.}}
  \bibinfo{year}{2011}\natexlab{}.
\newblock \showarticletitle{CrowdDB: answering queries with crowdsourcing}. In
  \bibinfo{booktitle}{\emph{{SIGMOD}}}.
\newblock


\bibitem[\protect\citeauthoryear{et~al.}{et~al.}{2013}]%
        {data-tamer}
\bibfield{author}{\bibinfo{person}{Michael~Stonebraker et al.}}
  \bibinfo{year}{2013}\natexlab{}.
\newblock \showarticletitle{Data Curation at Scale: The Data Tamer System}. In
  \bibinfo{booktitle}{\emph{{CIDR}}}.
\newblock


\bibitem[\protect\citeauthoryear{et~al.}{et~al.}{2023c}]%
        {prompt}
\bibfield{author}{\bibinfo{person}{Pengfei~Liu et al.}}
  \bibinfo{year}{2023}\natexlab{c}.
\newblock \showarticletitle{Pre-train, Prompt, and Predict: {A} Systematic
  Survey of Prompting Methods in Natural Language Processing}.
\newblock \bibinfo{journal}{\emph{{ACM} Comput. Surv.}} (\bibinfo{year}{2023}).
\newblock


\bibitem[\protect\citeauthoryear{et~al.}{et~al.}{2020a}]%
        {gpt3}
\bibfield{author}{\bibinfo{person}{Tom B.~Brown et al.}}
  \bibinfo{year}{2020}\natexlab{a}.
\newblock \showarticletitle{Language Models are Few-Shot Learners}. In
  \bibinfo{booktitle}{\emph{{NeurIPS}}}.
\newblock


\bibitem[\protect\citeauthoryear{et~al.}{et~al.}{2017}]%
        {holoclean}
\bibfield{author}{\bibinfo{person}{Theodoros~Rekatsinas et al.}}
  \bibinfo{year}{2017}\natexlab{}.
\newblock \showarticletitle{HoloClean: Holistic Data Repairs with Probabilistic
  Inference}.
\newblock \bibinfo{journal}{\emph{{VLDB}}} (\bibinfo{year}{2017}).
\newblock


\bibitem[\protect\citeauthoryear{et~al.}{et~al.}{2023d}]%
        {toolformer}
\bibfield{author}{\bibinfo{person}{Timo~Schick et al.}}
  \bibinfo{year}{2023}\natexlab{d}.
\newblock \showarticletitle{Toolformer: Language Models Can Teach Themselves to
  Use Tools}.
\newblock  (\bibinfo{year}{2023}).
\newblock


\bibitem[\protect\citeauthoryear{et~al.}{et~al.}{2020b}]%
        {ditto}
\bibfield{author}{\bibinfo{person}{Yuliang~Li et al.}}
  \bibinfo{year}{2020}\natexlab{b}.
\newblock \showarticletitle{Deep Entity Matching with Pre-Trained Language
  Models}.
\newblock \bibinfo{journal}{\emph{{VLDB}}} (\bibinfo{year}{2020}).
\newblock


\bibitem[\protect\citeauthoryear{et~al.}{et~al.}{2021b}]%
        {imp}
\bibfield{author}{\bibinfo{person}{Yinan~Mei et al.}}
  \bibinfo{year}{2021}\natexlab{b}.
\newblock \showarticletitle{Capturing Semantics for Imputation with Pre-trained
  Language Models}. In \bibinfo{booktitle}{\emph{{ICDE}}}.
\newblock


\bibitem[\protect\citeauthoryear{Freitas and Curry}{Freitas and Curry}{2016}]%
        {bigdc}
\bibfield{author}{\bibinfo{person}{Andr{\'e} Freitas} {and}
  \bibinfo{person}{Edward Curry}.} \bibinfo{year}{2016}\natexlab{}.
\newblock \showarticletitle{Big data curation}.
\newblock  (\bibinfo{year}{2016}).
\newblock


\bibitem[\protect\citeauthoryear{Izacard and Grave}{Izacard and Grave}{2021}]%
        {distillation}
\bibfield{author}{\bibinfo{person}{Gautier Izacard} {and}
  \bibinfo{person}{Edouard Grave}.} \bibinfo{year}{2021}\natexlab{}.
\newblock \showarticletitle{Distilling Knowledge from Reader to Retriever for
  Question Answering}. In \bibinfo{booktitle}{\emph{{ICLR}}}.
\newblock


\bibitem[\protect\citeauthoryear{OpenAI}{OpenAI}{2023}]%
        {chatgpt}
\bibfield{author}{\bibinfo{person}{OpenAI}.} \bibinfo{year}{2023}\natexlab{}.
\newblock \bibinfo{title}{ChatGPT}.
\newblock
\newblock


\bibitem[\protect\citeauthoryear{Schick and Sch{\"{u}}tze}{Schick and
  Sch{\"{u}}tze}{2021}]%
        {bootstrapping2}
\bibfield{author}{\bibinfo{person}{Timo Schick} {and} \bibinfo{person}{Hinrich
  Sch{\"{u}}tze}.} \bibinfo{year}{2021}\natexlab{}.
\newblock \showarticletitle{Exploiting Cloze-Questions for Few-Shot Text
  Classification and Natural Language Inference}. In
  \bibinfo{booktitle}{\emph{{EACL}}}.
\newblock


\bibitem[\protect\citeauthoryear{Xu and Zhu}{Xu and Zhu}{2022}]%
        {nci}
\bibfield{author}{\bibinfo{person}{Yichen Xu} {and} \bibinfo{person}{Yanqiao
  Zhu}.} \bibinfo{year}{2022}\natexlab{}.
\newblock \showarticletitle{A Survey on Pretrained Language Models for Neural
  Code Intelligence}.
\newblock  (\bibinfo{year}{2022}).
\newblock


\bibitem[\protect\citeauthoryear{Yarowsky}{Yarowsky}{1995}]%
        {bootstrapping}
\bibfield{author}{\bibinfo{person}{David Yarowsky}.}
  \bibinfo{year}{1995}\natexlab{}.
\newblock \showarticletitle{Unsupervised Word Sense Disambiguation Rivaling
  Supervised Methods}. In \bibinfo{booktitle}{\emph{{ACL}}}.
\newblock


\end{thebibliography}

\end{document}